\pdfoutput=1

\documentclass[sigconf]{acmart}

\AtBeginDocument{%
  }

\setcopyright{acmcopyright}

\copyrightyear{2023} 
\acmYear{2023} 
\setcopyright{rightsretained} 
\acmConference[CIKM '23]{Proceedings of the 32nd ACM International Conference on Information and Knowledge Management}{October 21--25, 2023}{Birmingham, United Kingdom}
\acmBooktitle{Proceedings of the 32nd ACM International Conference on Information and Knowledge Management (CIKM '23), October 21--25, 2023, Birmingham, United Kingdom}\acmDOI{10.1145/3583780.3614736}
\acmISBN{979-8-4007-0124-5/23/10}

\makeatletter
\gdef\@copyrightpermission{
 \begin{minipage}{0.3\columnwidth}
  \href{https://creativecommons.org/licenses/by/4.0/}{\includegraphics[width=0.90\textwidth]{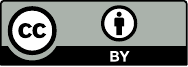}}
 \end{minipage}\hfill
 \begin{minipage}{0.7\columnwidth}
  \href{https://creativecommons.org/licenses/by/4.0/}{This work is licensed under a Creative Commons Attribution International 4.0 License.}
 \end{minipage}
 \vspace{5pt}
}
\makeatother

\usepackage{graphicx}
\usepackage{booktabs}
\usepackage[normalem]{ulem}
\usepackage{enumitem}

\newcommand{\toolname}{\textsc{CRUISE}--\emph{Screening}}

\usepackage[utf8]{inputenc}

\usepackage{soul} %
\usepackage{tikz}

\newcommand\encircle[1]{%
  \tikz[baseline=(char.base)] 
    \node (X) [draw, shape=circle, scale=0.7, inner sep=1pt, fill=black, text=white] (char) {#1};%
}

\title{CRUISE-Screening: Exploratory Search System For Academic Literature Reviews}
\title{CRUISE-Screening: Search System For Living Literature Reviews}
\title{CRUISE-Screening: Search System for Living Literature Reviews for Academics}
\title{CRUISE-Screening: Living Literature Reviews Toolbox for Academics}
\title{\toolname: Living Literature Reviews Toolbox}

\author{Wojciech Kusa}
\orcid{0000-0003-4420-4147}
\affiliation{%
  \institution{TU Wien}
  \city{Vienna}
\country{Austria}
  }
\email{wojciech.kusa@tuwien.ac.at}

\author{Petr Knoth}
\orcid{0000-0003-1161-7359}
\affiliation{%
  \institution{The Open University}
  \city{Milton Keynes}
\country{UK}
}
\email{petr.knoth@open.ac.uk}

\author{Allan Hanbury}
\orcid{0000-0002-7149-5843}
\affiliation{%
 \institution{TU Wien}
  \city{Vienna}
 \country{Austria}
}
\email{allan.hanbury@tuwien.ac.at}

\begin{document}

\begin{abstract}
Keeping up with research and finding related work is still a time-consuming task for academics.
Researchers sift through thousands of studies to identify a few relevant ones. 
Automation techniques can help by increasing the efficiency and effectiveness of this task.
To this end, we developed \toolname, a web-based application for conducting living literature reviews -- a type of literature review that is continuously updated to reflect the latest research in a particular field. 
\toolname~is connected to several search engines via an API, which allows for updating the search results periodically.
Moreover, it can facilitate the process of screening for relevant publications by using text classification and question answering models.
\toolname~can be used both by researchers conducting literature reviews and by those working on automating the citation screening process to validate their algorithms.
The application is open-source,\footnote{\url{https://github.com/ProjectDoSSIER/cruise-screening}} and a demo is available under this URL: \url{https://citation-screening.ec.tuwien.ac.at}.
We discuss the limitations of our tool in Appendix~\ref{sec:appendix}.
\end{abstract}



\maketitle

\section{Introduction}

Literature reviews are a critical component of all research domains.
Two different types of reviewing the literature can be distinguished: systematic reviews and, more general, exploratory literature reviews.
Systematic literature reviews follow strict criteria and are commonly used in healthcare and medical domains, as they provide the gold standard in evidence-based medicine~\cite{cook1997systematic,murad2016new}. 
Such reviews are a tedious, recall-oriented and repetitive process; they typically involve several stages, including formulating a research question, defining inclusion and exclusion criteria, searching multiple databases, screening for relevant studies, assessing study quality, extracting data, and synthesizing the findings. 
In recent years, several tools and procedures have been developed to automate this process, making it more efficient and effective.

The process applied in systematic literature reviews in medicine was also transferred into other scientific domains, such as environmental sciences~\cite{bilotta2014use}, software engineering~\cite{keele2007guidelines}, social sciences~\cite{petticrew2008systematic} and engineering~\cite{castillo2022apisser}.
There is an opportunity to reduce the human labour as well as the time it takes to deliver systematic reviews with the use of technology, specifically NLP and ML.

General literature reviews (usually more exploratory than ``systematic'') are also conducted in the academic setting, often by PhD and Masters students~\cite{green2009american,soufan2022searching}.
This process enables researchers to familiarise themselves with the current state of the art, theory and methods in their field.
Overlapping reviews are very often repeated by different groups as there is no data sharing and exchange format that could enable reusing past reviews as it is in systematic reviews in medicine~\cite{siontis2013overlapping,ioannidis2016mass}.
Guidelines and methodologies also aim to improve this process but do not mention any automated approaches, and the search process itself is not very structured but, instead, exploratory~\cite{pickering2014benefits}.
There is substantial potential in developing standards and tools that academics could adopt for the purpose of literature reviews.

In recent years, there has been a growing interest in the concept of living literature reviews~\cite{Elliott2014LivingGap,wijkstra2021living}. 
These reviews are continuously updated to reflect the latest research in a particular field. 

In this paper, we present \toolname, a web-based tool for conducting living literature reviews.
\toolname, developed to improve the efficiency of the literature review process, is connected to several search engines via API and facilitates the process of screening for relevant publications using NLP and machine learning methods.
We discuss the development and functionalities of \toolname, as well as present the challenges in developing such a tool. 
The system has notable novelty as it integrates search and screening capabilities into a single application and can connect with several machine learning models.
We foresee two use cases for our system: (1) primarily by researchers wanting to review the literature to locate the relevant work in their field of expertise; and (2) people developing automation models for literature reviews wanting to compare their approaches with others.

While most tools in this domain are developed specifically for systematic reviews, our system is among the first to apply systematic review concepts to general literature reviews. 
This sets our system apart from the rest of literature review tools, which are primarily recommendation systems for papers.
Our system has the potential to promote collaboration and facilitate the exchange of ideas among researchers.

\section{Related Work}

\subsection{Academic Search Engines}

Private academic search engines, citation indices and paywalled collections such as ScienceDirect and Web of Science are one source of finding publications.
Public search engines and publication aggregators such as Google Scholar\footnote{\url{https://www.scholar.google.com/}}, Semantic Scholar~\citep{ammar2018construction}, CORE~\citep{knoth2012core}, OpenAlex~\citep{priem2022openalex} and PubMed\footnote{\url{https://www.ncbi.nlm.nih.gov/pubmed/}} are becoming increasingly popular for allowing researchers to freely access the latest publications.
Their main goal is creating a citation network, and their support for conducting systematic literature reviews is often minimal.
Moreover, only a few of these tools provide an API, and none of them allow for a traditional systematic literature review workflow.

\subsection{Systematic Review Toolboxes}

There is already a number of tools helping researchers conduct systematic literature reviews.
An online catalogue\footnote{\url{http://www.systematicreviewtools.com}} enumerates 45 tools helping users during the screening phase, whereas~\citet{harrison2020software} found 15 of them accessible and available without specific computing infrastructure for a title and abstract screening step.
Dedicated commercial tools exist to help medical researchers conduct systematic literature reviews.
They are usually customised to medical reviews and require purchasing a subscription which can be a bottleneck to academic researchers from lower- and lower-middle income countries~\cite{morales2022resource,nussbaumer2022resource}.

In addition to the commercial tools, a plethora of free or open-source tools is available, usually created by academics.
These tools, such as Abstrakr~\cite{Wallace2012DeployingAbstrackr}, Rayyan~\cite{elmagarmid2014rayyan}, or ASReview~\cite{van2021open} usually support only one of the systematic review stages.

\subsection{Automated Citation Screening}

All the documents retrieved from the search step constitute the input to the citation screening step.
In a manual screening scenario, reviewers read all these documents to select only the fraction relevant to the systematic review. %
Because the total number of retrieved studies can go into tens of thousands, it is essential to find a way of improving this process~\cite{OMara-Eves2015}.
Automated citation screening is an umbrella term for using NLP, machine learning and information retrieval (IR) techniques with the goal of decreasing the time spent on manual screening.
Classification approaches train a supervised model on an annotated dataset to determine whether a paper should be included or excluded from the review~\cite{Kusa2023Analysis,kusa2023vombat}.

\begin{figure*}[ht]
    \centering
    \includegraphics[width=0.87\textwidth]{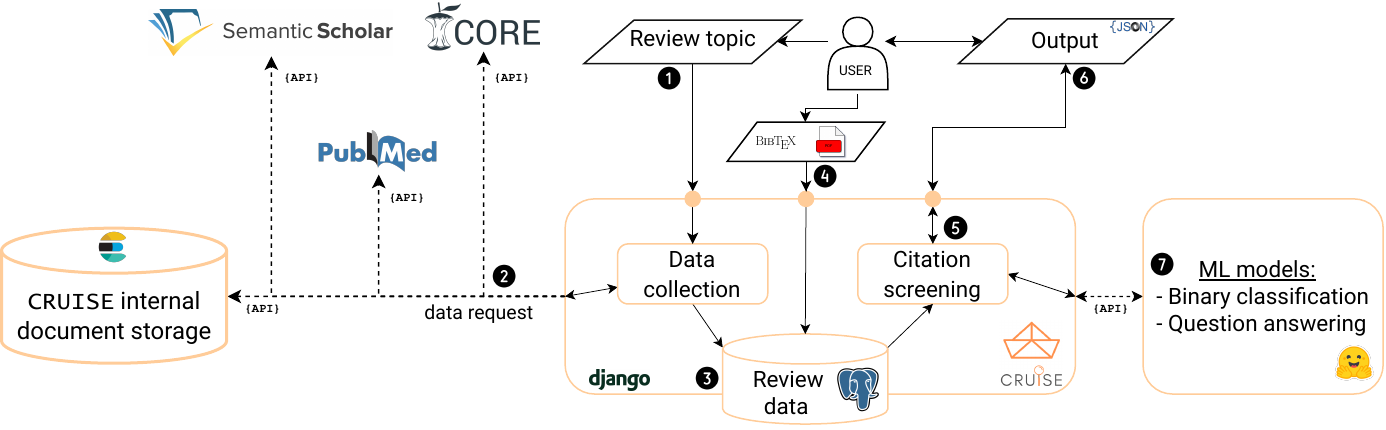}
    \caption{Overview of the \toolname~architecture.}
    \label{fig:cruise_architecture}
\end{figure*}

Previous approaches ranged from statistical models like naïve Bayes classification~\cite{bekhuis2014feature,Matwin2010AReviews}, support vector machine (SVM)~\cite{Cohen2011Letter:Measure,martinez2008facilitating,Howard2016a,Cohen2008OptimizingPrioritization}, voting perceptron~\cite{Cohen2006} and random forest~\cite{Khabsa2016LearningInformation} to neural networks~\cite{Kontonatsios2020,Kusa2022AutomationStudy}.
A significant limitation of all these approaches is the need for a large number of manual annotations that must be completed before developing a reliable model for every new systematic review~\cite{Tsafnat2018AutomatedCharacteristics}.
Moreover, the majority of the classification models are evaluated only retrospectively which might raise questions of data leakage when considering large amounts of data used for pretraining language models~\cite{kusa2023outcomebased}.

\section{CRUISE-Screening}

Figure~\ref{fig:cruise_architecture} shows the architecture of \toolname, which is built with Python~3.9, Django~4, Bulma and AlpineJS frameworks.

\subsection{Data Resources} \label{sec:data}

Good quality input data covering multiple domains is the crucial ingredient of a successful literature review.
\citet{NUSSBAUMERSTREIT20181} found that combining two separate databases may suffice to reliably determine the conclusions of a systematic review in medicine.
Therefore, \toolname~was designed to use multiple data sources and to allow for extending them when needed.
Currently, it supports the following four search engines as data sources: Semantic Scholar API\footnote{\url{https://www.semanticscholar.org/product/api}}, CORE API\footnote{\url{https://core.ac.uk/services/api}}, PubMed via \textsc{Entrez} API\footnote{\url{https://www.ncbi.nlm.nih.gov/search/}} and internal document storage.

The first three APIs call search engines that are used as primary data sources when searching for documents.  
Using three different search engines with contrasting scopes and content enables good search results coverage. 

The tool also allows for indexing documents in the internal database. 
It is implemented using Elasticsearch and communicates with the main application using the API.
It can be used, for example, when one wants to store private documents or content not covered by other search engines. 
For this demo, we index the DBLP-Citation-network Version 13 collection\footnote{\url{https://www.aminer.cn/citation}} created by~\citet{Tang:08KDD}.

The system could be expanded to connect to other search engines offering API access.
As the system is a meta-search engine, we use a script to deduplicate the search results based on papers' metadata.

\subsection{Screening Workflow}

The typical screening workflow for systematic literature reviews consists of two stages. 
In the first stage, the researcher searches for documents potentially related to the research topic.
In the second stage, the documents are screened for relevance. 
We have implemented these two stages inside \toolname.

\subsubsection*{Search for relevant items}

\begin{figure}[tb]
    \centering
    \includegraphics[width=0.94\columnwidth]{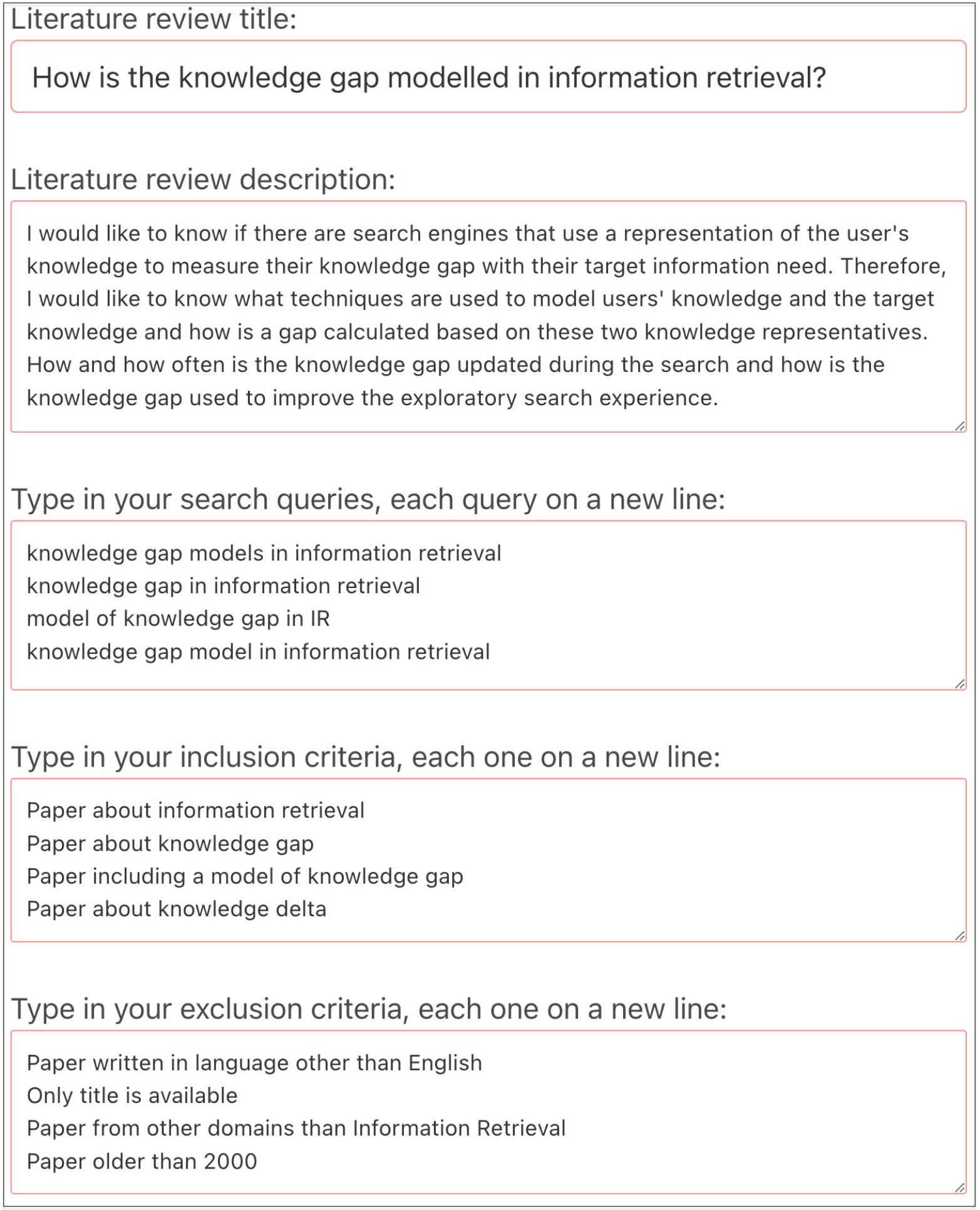}
    \caption{Example literature review protocol containing review title, description, search queries and criteria for inclusion and exclusion.}
    \label{fig:review_protocol}
\end{figure}

First, the user creates a new literature review by defining the research protocol~\encircle{1}.
The protocol (Figure \ref{fig:review_protocol}) consists of the review's title, description, at least one search query and a set of inclusion and exclusion criteria (eligibility criteria).
The tool allows for specifying search engines, by default selecting all four available sources described in Section \ref{sec:data}.
The search can be limited to only the first $N$ results if the reviewer is not interested in a comprehensive literature review.

\toolname~sends API requests to selected search engines and gathers all responses~\encircle{2}.
Merged and deduplicated search results are stored in a PostgreSQL database~\encircle{3}.
In order to support living reviews, the user can re-run the search function periodically to update the list of references.
However, since search engines only allow filtering by publication year and not month or day, the tool removes publications older than the year of the previous search during updates. 
The tool then relies on deduplication to ensure that new publications are not mistakenly added twice.

\begin{figure*}[ht]
    \centering
    \includegraphics[width=0.84\textwidth]{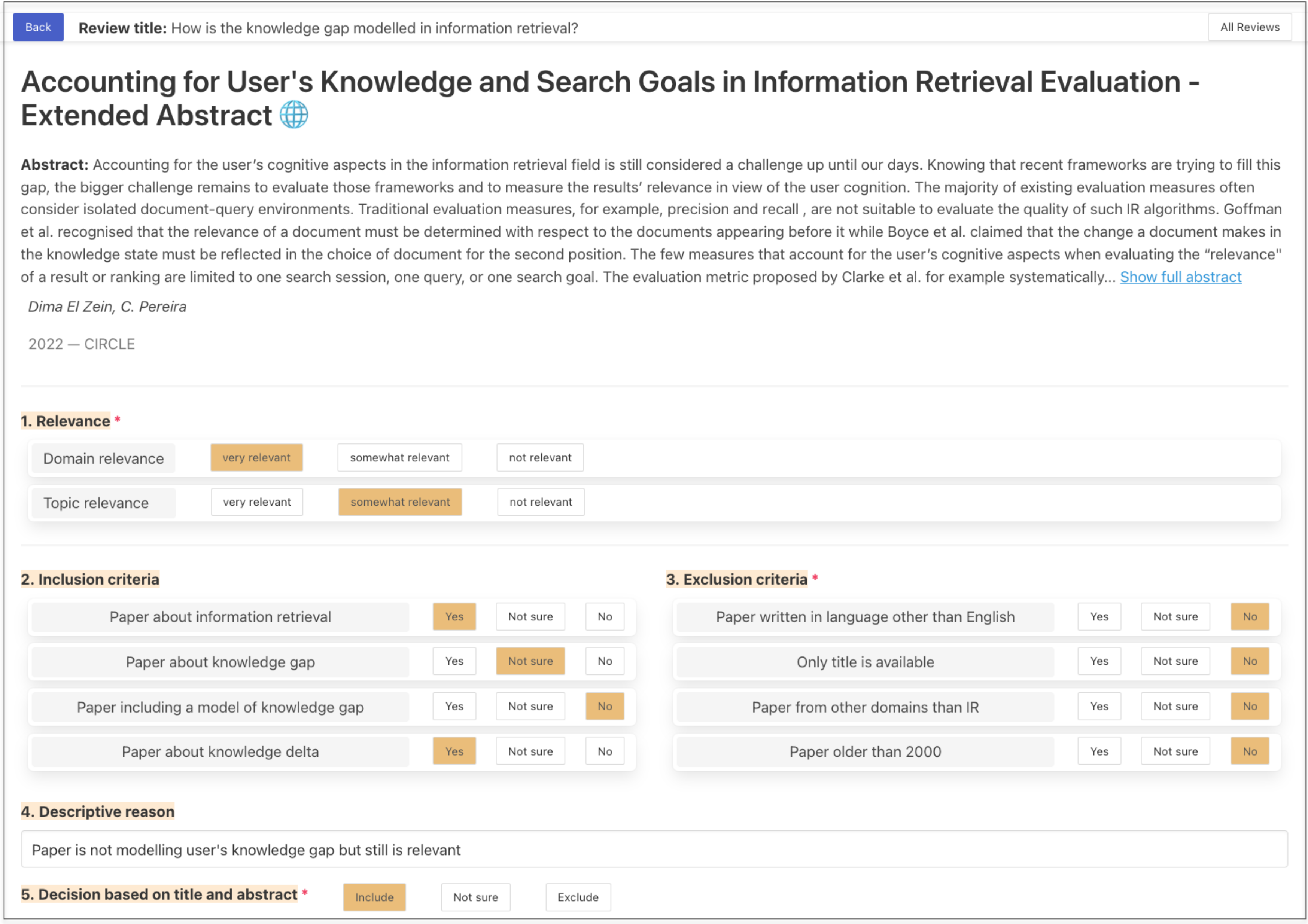}
    \caption{Example screening interface in \toolname~presenting single paper with answered questions.}
    \label{fig:cruise_interface}
\end{figure*}
\toolname~also allows for the additional direct import of data for screening from two sources~\encircle{4}: 
\begin{itemize}[leftmargin=14pt]
    \item Bulk upload from reference files -- Currently, the tool supports \textsc{bib} and \textsc{ris} file extensions. %
    These publications are imported to the PostgreSQL database.
    \item Full text \textsc{pdf} files -- %
    These files are processed using \textsc{GROBID}~\citep{GROBID} and then added to the database. Documents added this way can also be marked as seed studies. This way, these new documents are labelled as \textit{relevant}, which can speed up the process of automated screening.
\end{itemize}

\subsubsection*{Citation screening}

Currently, \toolname~implements the title and abstract screening~\encircle{5} while providing external URLs to full text articles whenever available.
Figure \ref{fig:cruise_interface} presents an example screening interface.
From the top, it contains the title, abstract, authors, publication venue and year and the link to the full text of the screened paper.
Below are two sections with eligibility criteria questions and a main inclusion question.

There are two screening workflows in \toolname: strict and relaxed.
Strict screening requires the annotators to conduct the process by manually answering every eligibility question.
It mimics the citation screening process of systematic reviews.
This mode could be used for in-depth systematic reviews or gathering manual annotations for training machine learning models.

The relaxed mode does not impose any requirements on which questions the annotator should answer except for the main \textit{include}/ \textit{maybe}/\textit{exclude} decision.
There are optional questions about the reviewer's prior knowledge of the paper and authors, which reviewers can turn on to control for the selection bias.

The output of the literature review can be exported in a \textsc{json} format~\encircle{6}.
It contains the literature review protocol and all identified studies with corresponding automatic and manual decisions.

\subsection{Automation Methods}

Except for the fully manual workflow, \toolname~implements automation methods to increase the speed and coverage of the literature review~\encircle{7}.
Implemented approaches include supervised text classification and zero-shot question-answering models.
The tool connects to them using an API, which allows for extending the list of supported models.

\subsubsection*{Text classification}

We implemented two examples of supervised classifiers based on previous literature: a logistic regression model using tf-idf text representation and a fastText classifier~\cite{Joulin2016}.
These models provide a single \textit{yes}/\textit{no} decision for each paper (corresponding to the main eligibility question from the manual workflow).
Reviewers need to annotate a ``training set'' of at least three included and three excluded papers before using the models.
When the reviewer annotates more publications, the models can be retrained to make an updated prediction on the remaining documents.

\subsubsection*{Question answering}

In addition to supervised text classification, \toolname~enables users to conduct automatic screening using prompt-based language models with a question answering approach.
The advantage of this method is that it does not require pre-labelled data and can make predictions for all inclusion and exclusion criteria.
However, it can be computationally intensive and sensitive to the quality of input questions.
The API is designed to support any Text2TextGeneration model implemented in the HuggingFace Transformers~\cite{wolf2019huggingface} library.
We used the T0\_3B and T0 models~\cite{sanh2021multitask}.
The example prompt consists of a single eligibility question and the same paper data as available during manual screening (Figure~\ref{fig:cruise_interface}), namely the title, abstract, authors, journal name and publication year.

\section{Discussion and Conclusion}

The merging of results from multiple sources can present significant challenges in the context of scientific publications. 
Although using multiple data sources can lead to better coverage of relevant studies, combining the results from different sources is not a trivial task. 
The data can have different formats, fields, and identifiers, which require significant effort to reconcile.
Additionally, data quality can be poor in some cases, which can further complicate the merging process. 
Therefore, careful consideration must be given to the merging process, and the use of automated tools can help improve the accuracy and efficiency of the process. 
Nevertheless, human intervention may still be required to resolve any inconsistencies or errors in the data~\cite{guimaraes2022deduplicating}.

The evaluation of models is an essential aspect of our tool, as it allows researchers to evaluate their models without the risk of data leakage. 
Due to the vast amount of training data used by large language models, it can be difficult to detect data leakage when retrospectively evaluating on common benchmarks. 
Therefore, our tool provides a solution by enabling researchers to make predictions with several models before starting a new review, storing the results, and evaluating the models after the manual review is conducted, without interfering with the manual workflow.

Compared to other tools, \toolname~combines search and screening stages into one workflow.
Thanks to this, researchers can use the tool as an information system to systematise,  manage and record their literature review workflows.

We note that the approach based on prompting large language models can generate non-reliable predictions.
We added warnings to the user interface so the user knows these predictions can contain hallucinations.
In future work, we plan to present the user with a predicted performance on this particular criterion if the same question was asked in the previous reviews and there was enough data on its accuracy against the ground truth.

\section*{Acknowledgements}
This work was supported by the EU Horizon 2020 ITN/ETN on Domain Specific Systems for Information Extraction and Retrieval -- DoSSIER (H2020-EU.1.3.1., ID: 860721).
We thank all the members of the DoSSIER Project who contributed to the CRUISE workshop.

\bibliography{anthology,custom}
\bibliographystyle{acl_natbib}

\appendix

\section{Limitations}
\label{sec:appendix}

This section discusses the limitations that should be considered when using \toolname.

\begin{description}[leftmargin=14pt]

\item \textbf{Data sources:} The \toolname~relies on APIs to conduct the search as it acts as a meta-search engine. 
These APIs could disappear or change over time, affecting the tool's functionality. 
However, given that we are using multiple resources at once, the risk of this limitation should be mitigated.
Moreover, although \toolname~is connected to several search engines, it may not cover all potential databases or specialised repositories, potentially missing out on some relevant literature.

\item \textbf{Search technique:} We do not rely on Boolean queries but a set of keyword-based queries, which together, create a pool of retrieved documents. This approach differs from classic systematic reviews. Additionally, we limit the search to the top 500 records for each query by default to speed up the process, which could potentially limit the coverage of relevant studies.

\item \textbf{Hallucinations:} Large language models can sometimes ``hallucinate'' and create incorrect predictions or outputs. 
Users should be aware that the automated screening process could produce false positives or false negatives due to these hallucinations.

\item \textbf{Biases:} The machine learning models used for screening could have biases in their predictions due to biased training data, which could impact the quality and representativeness of the literature review.

\item \textbf{Cost:} The deployment and continued use of large language models in the \toolname~can be expensive. 
The computational requirements for training and deploying these models are substantial, and as models grow in size and complexity, the associated costs may increase. 
This could potentially impact the scalability and affordability of the tool for researchers with limited resources or budget constraints.

\item \textbf{User experience and accessibility:} While \toolname~is designed to be user-friendly, there might be a lack of sufficient detail on accessibility features, potentially making it challenging for a wider range of researchers, including those with specific needs or non-technical backgrounds, to use the tool effectively. 
Furthermore, the current design of the user interface, while functional, may not be optimal for all potential users. 
We recognise the need for further user studies to assess its intuitive nature and to identify areas of improvement. 
We aim to improve the tool's usability and accessibility in future iterations.

\end{description}

\end{document}